\documentclass[preprint,12pt,authoryear]{elsarticle}

\usepackage{amssymb}
\usepackage{amsmath}

\usepackage[colorlinks=true]{hyperref}

\usepackage{xspace}
\newcommand{\cotwo}{CO$_2$\xspace}

\usepackage{xcolor}
\usepackage{mdframed}

\newmdenv[
  backgroundcolor=gray!15,
  linecolor=gray!50,
  linewidth=0.5pt,
  skipabove=\topsep,
  skipbelow=\topsep
]{grayquote}

\usepackage{booktabs}

\journal{Journal of Systems and Software}

\begin{document}

\begin{frontmatter}



\title{Energy-Aware Decision Making in Software Stack Upgrades}



\author{Mirko Stocker}

\affiliation{organization={Eastern Switzerland University of Applied Sciences (OST)},
            addressline={Oberseestrasse~10}, 
            city={Rapperswil},
            postcode={8640}, 
            state={St.~Gallen},
            country={Switzerland}}
            
\author{Michael Wahler}

\affiliation{organization={Zurich University of Applied Sciences (ZHAW)},
            addressline={Obere Kirchgasse~2}, 
            city={Winterthur},
            postcode={8400}, 
            state={Zurich},
            country={Switzerland}}

\begin{abstract}

    
Software stack upgrades are a routine part of software maintenance and evolution, typically motivated by improved performance, stability, or functionality. Yet their impact on \emph{energy consumption}---a growing concern for organizations pursuing sustainability---remains poorly understood. This paper presents a systematic method for measuring how different versions of core software stack components, such as Spring Boot and the Java Virtual Machine (JVM), influence the energy consumption of applications. Our approach evaluates combinations of framework versions, runtime versions, and execution platforms through automated benchmarking.
Using a case study based on the Spring Petclinic REST application, we show that energy consumption varies substantially across Spring Boot and JVM versions, in some cases producing unexpected regressions. Notably, newer JVM releases and virtual threads (introduced in Java 21 and 23) yielded significant energy savings without requiring application changes.

These results demonstrate that software upgrades can meaningfully affect energy usage and that measuring energy consumption provides valuable evidence for decision making in software maintenance and evolution.
\end{abstract}


\end{frontmatter}

\section{Introduction}\label{introduction}
Organizations routinely upgrade components of their software stacks in software maintenance and evolution~\citep{lehman_programs_1980}, and different reasons for performing maintenance activities have been identified~\citep{swanson_dimensions_1976}.
However, despite growing sustainability awareness, the impact of such upgrades on energy consumption remains largely unexplored. This gap is increasingly important as software (particularly, cloud‑hosted systems) accounts for a non‑trivial and rising share of global energy usage. 
The predominant reasons for software upgrades are maintainability and performance~\citep{wohlin_towards_2021}, which are aspects of the technical sustainability~\citep{McGuire:2023} of software systems. 
Recent research has started to investigate energy consumption as an additional factor in software upgrades, especially in mobile and embedded systems~\citep{sas_quality_2020}.
For general software systems, 
energy considerations play a negligible role in decision making for software upgrades today.
While prior research has examined energy efficiency in programming languages, architectures, and configurations, little is known about how upgrading widely used frameworks and runtimes affects the energy profile of real applications.

\subsection{Sustainability in IT}

Energy consumption of IT systems has become an important topic in the context of climate change because of the \cotwo emissions of the fossil fuels used for energy production.
There are two ways of looking at software and sustainability: \emph{Sustainability by IT} and \emph{Sustainability in IT} \citep{Calero:2021}. Sustainability by IT means that software can act as an enabler with which sustainability gains can be achieved, such as scheduling online meetings instead of flying to in-person meetings. But there is also a flip side to the increased use of software, so we also have to consider the sustainability in IT itself.

A study by \citet{Freitag:2021} estimates that the share of the Information Communication Technology (ICT) sector could already be as high as 3.9\,\% of total global emissions. These calculations cover the entire hardware lifecycle, from manufacturing and operations to disposal, and include components such as data centers, networks, and end-user devices. During operation, the software has a significant influence on the amount of energy that is used. So, in addition to using electricity from sustainable sources, one step towards sustainability is to minimize the energy consumption of the underlying hardware caused by software, including processing (CPU, RAM), storage, and network devices.

\newcommand{\CC}{C\nolinebreak\hspace{-.05em}\raisebox{.4ex}{\tiny\bf +}\nolinebreak\hspace{-.10em}\raisebox{.4ex}{\tiny\bf +}\xspace}

On the software side, there are different approaches to optimizing the energy consumption: programming languages differ in their energy consumption, cloud software architectures can also have an impact \citep{Procaccianti:2013}. By moving to the cloud or using virtualization, existing hardware can be better utilized, and different strategies can be applied to improve energy efficiency \citep{Procaccianti:2015}. For example, studies on the energy efficiency of programming languages have shown that choice of program language can have a significant impact on the energy consumption of software \citep{Gordillo:2024}. However, can we expect developers to ditch the programming language they are most productive in, their favorite frameworks that they know inside out and rewrite all their software in C or \CC for something as abstract as energy consumption or sustainability? Probably not, at least not in the short term. But there might be quick wins, like upgrading to the latest versions of third-party components---assuming that these newer versions are more energy-efficient. But how can we know if this is actually the case?

\subsection{Contributions}\label{contributions}

\begin{figure}
    \centering
    \includegraphics[width=\linewidth]{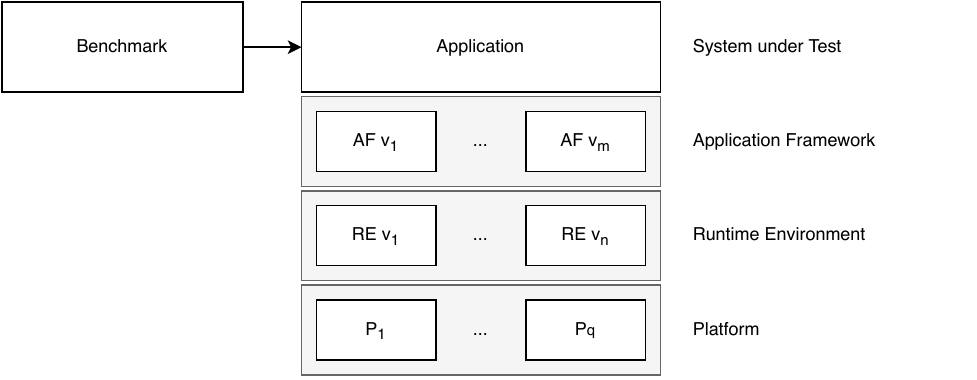}
    \caption{Software and hardware stack for our energy measurement approach.}
    \label{fig:architecture-generic}
\end{figure}

In this paper, we present a systematic approach for measuring the energy consumption of an application across different versions of the underlying software and hardware stack. As illustrated in \autoref{fig:architecture-generic}, the application serves as the system under test, while the stack below it---comprising the application framework, runtime environment, and platform (hardware and operating system)---is varied in a controlled manner. Our method evaluates all combinations of the available $m$ framework versions, $n$ runtime versions, and $q$ platforms, allowing us to quantify how each component influences the application's total energy usage. This is achieved by automatically deploying each possible combination of versions and performing a benchmark, which creates the same workload for each such deployment.

The resulting measurements provide an evidence base that can support or challenge upgrade decisions in software maintenance and evolution. We propose that these energy measurements complement traditional upgrade drivers such as performance improvements, bug fixes, and new features, enabling developers and organizations to incorporate energy consumption as an additional decision criterion. 

We demonstrate our approach with a case study using the Spring Web application framework and the Java Virtual Machine (JVM). Spring is a popular framework for building Web applications in Java and other programming languages that run on the JVM. According to the JetBrains Developer Ecosystem survey \citep{Jetbrains:2023}, Spring is used by a large majority of Java developers to develop backend applications, often APIs, running on servers in private data centers or public clouds. According to the Spring Website \citep{SpringWebsite:2025}, Spring is ``everywhere, flexible, productive, fast, secure, and supportive''. But how energy-efficient is it? According to a study by \citet{Calero:2021:Spring}, applications ``developed using Spring need much more energy than those developed without using it.'' This comparison is based on the same application, implemented with and without Spring over the course of three releases and recommends to ``minimise the use of Spring in developing software wherever possible''. While some overhead from using a framework is expected, we believe this does not justify asking developers to abandon a framework in which they are productive and start over from scratch. Instead, we aim to investigate whether existing Spring applications can be improved in terms of energy efficiency.

Applications written using Spring run on the JVM, which executes the Java byte code generated by the Java compiler. Multiple versions of the JVM exist, and their version and specific configuration might also have an impact on the energy consumption of the software running on them. 
Furthermore, recent JVM versions support the concept of virtual threads, which promise better energy efficiency through a reduced overhead, especially for I/O intensive tasks.
Using this approach, we seek to answer the following overall research question: 

\begin{quote}
\textit{Can differences in energy consumption between components of a software stack be identified using automated measurements and component substitution?}
\end{quote}

To answer this research question, we developed a benchmark that simulates interactions with the Spring Petclinic API, an application provided by the Spring community that showcases Create, Read, Update, Delete (CRUD) operations that are common in many backend applications. Our case study covers the following aspects that can help inform upgrade decisions:

\begin{itemize}
  \item How does the energy consumption of the application vary across different versions of the Spring Boot framework?
  \item How does the energy consumption of the application vary across different versions of the JVM?
  \item Does the targeted Java compilation version of components influence the energy consumption?
  \item Do changes of execution paradigms/semantics in the JVM have an impact on the energy consumption?
\end{itemize}

To answer these questions, we measured the energy consumption of the benchmark running on different versions of the Spring framework and the JVM using JoularJX by \citet{Noureddine:2022}, a software-based tool for measuring the energy consumption of Java applications. 

The primary contributions of our research are:

\begin{itemize}
\item \textbf{A practical, reproducible measurement method} for quantifying how software stack upgrades affect application energy consumption.
\item \textbf{A fully automated benchmarking pipeline} that evaluates all combinations of framework and JVM versions, enabling evidence‑based upgrade decisions.
\item \textbf{A multi‑platform empirical evaluation}, showing that energy usage varies significantly across Spring Boot releases, JVM versions, and CPU architectures.
\item \textbf{Actionable findings for practitioners}, including:
    \begin{itemize}
        \item newer JVMs often reduce energy use,
        \item virtual threads provide measurable energy savings with no application changes, and
        \item compilation target version does not materially affect energy consumption.
    \end{itemize}
\end{itemize}

The remainder of this paper is structured as follows: \autoref{related-work} provides an overview of related work. \autoref{experiment-design} explains our methodology for the setup of the experiments, \autoref{results} presents the results of the experiments. \autoref{discussion} discusses the findings, \autoref{threats-to-validity} outlines threats to validity, \autoref{conclusion} concludes the paper, and \autoref{future-work} gives an outlook on future work.

\section{Related Work}\label{related-work}

Previous research has addressed decision making in software evolution and energy efficiency in software.

\subsection{Decision Making in Software Evolution}
In software maintenance and evolution, numerous decisions (and trade-offs between them) need to be made. 
Existing literature on this topic provides models and frameworks for guiding these decisions.

\cite{krishnan_decision_2004} have developed a decision model that provides a quantitative framework for deciding whether, how, and when to upgrade. This model compares the costs of ongoing minor upgrades and operational maintenance with the large, fixed costs of replacement based on factors such as the rate of system deterioration. To make decision models reusable, they should be captured \citep{Zimmermann:2007}, for example using Markdown \citep{Kopp:2018}.

Patterns of performance variations have been reported by \citet{Alcocer:2015}, which they mined from applications with 19 million lines of code. They have found that ``1 out of every 3 application revisions introduces a performance variation.'' They propose a measurement methodology structured around a nine steps workflow to identify performance variations over multiple software revisions.

A framework for evidence-based decision-making in software evolution, focusing on evaluating reusable components and balancing quality aspects such as maintainability and performance has been presented by~\cite{wohlin_towards_2021}. Their framework is applicable to arbitrary quality attributes (such as energy consumption, but it is not mentioned in the cited work). 

\cite{franco_systems_2023} have proposed systems thinking to model how long-term maintenance decisions influence technical debt and maintainability, offering a causal-effect map of architectural and maintenance trade-offs. Criteria such as size and quality of the software (and its requirements) are in the focus, whereas energy consumption is excluded.

Energy efficiency has been addressed by \cite{sas_quality_2020} in their study on how practitioners weigh trade-offs between maintainability, dependability, and energy efficiency. Their work emphasizes methods like ATAM (Architecture Trade-off Analysis Method) for evaluating architectural decisions. With a focus on embedded systems, which mostly use lean and simple software stacks, their results are difficult to apply to general-purpose computer systems.

\subsection{Energy Efficiency in Software}
Energy efficiency in software has been studied by many researchers from different perspectives. One point of view is from the programming language: \citet{Gordillo:2024} have ranked over a dozen programming languages by their energy efficiency when running different algorithms, using a hardware-based power meter. They have also compared their results to a previous study \citep{Pereira:2021} using a software-based approach and concluded that ``no significant differences are obtained in the [hardware-based] and [software-based] rankings.''

\citet{Pereira:2017} have studied 27 different implementations of programming languages for the correlation between execution time, memory consumption and energy efficiency. They have found that the faster language is not necessarily the more energy efficient one, due to differences in memory consumption. This relationship is further discussed, and the previous results criticized, by \citet{Kempen:2024}. They rightly point out the missing ``separation of concerns'' \citep{Dijkstra:1974}, that it is not the programming language (defining syntax and semantics), but the implementation that primarily influences performance and thus energy efficiency. While the difference between (abstract) language and (concrete) implementation may seem like a minor semantic inaccuracy, it does have an impact, as they show. For example, the degree of parallelism may vary between implementations and versions and should therefore also be considered.

If it is not feasible to change the programming language, the energy consumption of programs can be optimized by replacing parts of the code with semantically equivalent, but more efficient versions. This is accomplished by the SEEDS framework by \citet{manotas_seeds_2014} in which different variants of programs are created, tested, and their energy consumption measured. The authors evaluate their framework by optimizing the use of the \emph{Collections} package in Java programs, which offers different types of collections. Each collection type is optimized for specific use cases and therefore has a different computational complexity and energy footprint. Compared to our approach, SEEDS focuses on optimizing the application itself, whereas our approach focuses on optimizing the stack underneath the application (see \autoref{fig:architecture-generic}).
\cite{aggarwal_greenadvisor_2015} have investigated the changes in power consumption between different versions of system-level software. Their approach complements our approach by analyzing the code written in a project, whereas our approach focuses on components in the software stack that are used ``out of the box''. 

Not only the choice of programming language implementation has an impact on energy efficiency, but also the choice of application framework \citep{Calero:2021:Spring} or auxiliary frameworks. \citet{Bonvoisin:2024} have studied the influence of different object-relational mapping (ORM) frameworks and ``showed that energy efficiency might come at the price of performance and that developers may face potential trade-offs when configuring their ORM framework.'' The choice of ORM is just one decision among many that application architects and developers have to make. Databases, libraries for authentication, searching, caching, etc. are other decisions that must be made. JHipster is an application generator that offers many options for these, resulting in thousands of different combinations, each with a different performance profile, as \citet{Guegain2024} have shown. Not only the choice of framework or library matters, also how it is configured, which can also change between versions. \citet{Kaltenecker:2023} have found that ``almost every release of every subject [system] contains, at least, one performance change in some configuration'', and \citet{Weber:2023} have shown ``that the correlation between energy consumption and runtime performance indeed depends on individual configuration options and interactions thereof.'' On a smaller scale, the choice of data structure and design pattern also has an impact on energy consumption \citep{Pereira:2016,Noureddine:2025}.

Several studies have investigated the impact of different Java virtual machines on energy consumption. \citet{Ournani:2021} have selected several versions of JVM implementations and distributions (GraalVM, HotSpot, J9), and compared their energy efficiency using open-source benchmarks. Their results show that ``GraalVM reported the best energy efficiency for a majority of benchmarks'' and that the latest releases tend to be more energy efficient. As part of our study, we will see if this still holds for the currently available JVMs (the JVM versions 8--15, which they study, have all reached end of life by now). They have gone even further and also compared JVM parameters and configuration settings, finding that ``the default [just-in-time] compiler of the JVM is often near-optimal'', but that the ``default [garbage collector] consumed more energy than other strategies in most of the situations.'' \citet{Vergilio:2023} have also investigated energy efficiency of JVM variants and came to a similar conclusion: ``It is recommended that programmers and technology businesses consider adopting GraalVM in their future Java applications because of its energy efficiency.'' To the initiated, this result does not come as a surprise: the defining distinction of GraalVM among JVMs is that it does ahead-of-time compilation to produce a native binary. Unfortunately, GraalVM is not simply a drop-in replacement for most projects. Ahead-of-time compilation severely restricts runtime reflection and flexibility, features on which the Spring framework heavily relies.

Given all these factors that influence energy consumption of software systems, what is the developer's perspective? \citet{Pang:2016} have ``found that the programmers had limited knowledge of energy efficiency, lacked knowledge of the best practices to reduce software energy consumption, and were often unsure about how software consumes energy.'' \citet{Manotas:2016} have reported on an ``empirical study of how practitioners [who appeared to have experience with green software engineering] think about energy'' in the different development stages. Their main conclusion is that ``green software engineering practitioners care and think about energy when they build applications; however, they are not as successful as they could be because they lack the necessary information and support infrastructure.'' They also note that ``energy concerns are largely ignored'' in the maintenance stage and presume that by then ``it is either too difficult or not important enough to change energy usage''.
This is confirmed by a more recent study from \cite{10.1145/3639475.3640109}, who have found that the stakeholders involved in software development projects are typically not concerned with the energy usage of the software under development (e.g., by encouraging the use of efficient data structures) or the software development process itself (e.g., by optimizing the energy use of the CI/CD pipeline).

\subsection{Our Research}
With an ever-growing use of digital services, which become more powerful and power-hungry through the inclusion of AI, energy consumption is an important aspect of maintaining and evolving computer systems.
The research presented in this paper fills an important gap in the existing research.
By providing a method for measuring and transparently presenting the energy consumption of different software component versions, existing methods and frameworks for decision making can be augmented to include energy considerations.

\section{Methodology}\label{experiment-design}

In this section, we describe the design of our approach to measuring energy consumption in our case study, following the structure and process outlined in the Green Software Measurement Model (GSMM). The GSSM is a ``reference measurement model for assessing the resource and energy efficiency of software products and components'' \citep{Guldner:2024}. The model components are ``Measured Object and Measurement Goals'', ``Measurement and Metrics'', ``Measurement Procedure Models'', ``Measurement Setup'', and ``Data Evaluation Models''. This section is structured according to the GSMM. The artifacts are published on Zenodo for transparency and reproducibility \citep{ZenodoDataset:2025}.

\subsection{Measured Object and Measurement Goals}\label{measured-object-and-measurement-goals}

The measured object in our study is the energy consumption of software applications running on different versions of the Spring framework, the JVM, and hardware/OS platforms. Our goal is to understand how different versions of these two technologies affect the energy consumption of software applications. Our System under Test (SuT) is the Spring Petclinic REST sample application, which demonstrates the use of the Spring Framework within the sample domain of pet management, including veterinarians, owners, and their interactions. The API offers various Create, Retrieve, Update, and Delete (CRUD) operations, which are typical for API backend applications (\citet{Serbout:2022} have shown that CRUD and REST are the dominant styles in larger APIs). We chose Spring Petclinic because it demonstrates the idiomatic and best-practice use of technologies and is developed by several Spring maintainers in the official spring-projects repository. It uses several technologies that are commonly found in Spring applications, such as:

\begin{itemize}
    \item Spring Data Java Persistence API (JPA) for CRUD operations; 
    \item Spring Model-View-Controller (MVC) with components that handle calls to its API; and
    \item Dependency Injection (DI) and Inversion of Control (IoC), with the lifecycle of components being managed by the Spring container.
\end{itemize}

This makes Petclinic a suitable and reliable system under test for our study. 
Some features of Spring that could have an influence on the energy consumption of Spring applications are not present in Petclinic such as
\begin{itemize}
    \item microservices, which introduce communication overhead between system components;
    \item messaging and integration with other systems using Kafka or RabbitMQ, which creates additional I/O workload; and
    \item observability and monitoring, which also has an impact on I/O.
\end{itemize}

For applications that extensively use these features, energy measurements could yield different results than Petclinic. The same applies to applications that use computationally heavy business logic rather than mostly transferring and manipulating data. 
Our method can, however, be simply applied to such applications by re-running our measurements with a different application replacing Petclinic.

\begin{figure}[ht]
\centering
\includegraphics[width=.9\textwidth]{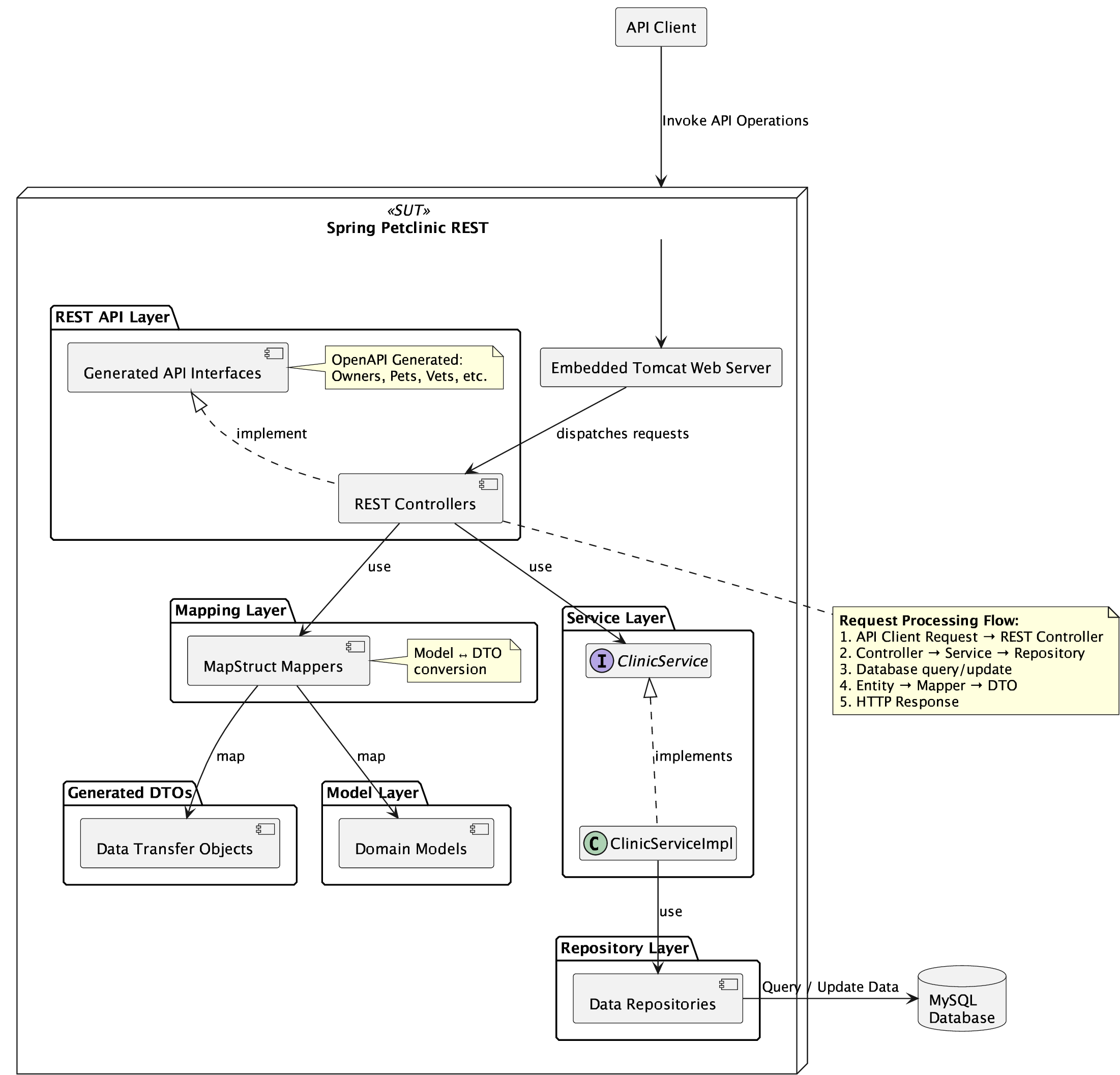}
\caption{Architectural overview of the example application used for the experiments in this paper}\label{fig:petclinic-architecture-diagram}
\end{figure}

The Petclinic application uses a relational database to store the application data. By default, an in-memory database is used, but we configured the application to use an external MySQL database, since the database's energy consumption is not relevant for this experiment; we are only interested in the example application, respectively the JVM it is running on. See \autoref{fig:petclinic-architecture-diagram} for an architectural overview of the Petclinic application.

A further benefit of using this sample application and Spring in general is that it runs with different framework and JVM versions without any or minimal changes to the source code. This allows us to vary the versions without having to touch the source code and distorting the measurements by using different implementations of the applications (researching the impact of different design decisions would be another interesting topic for further research, but is out of scope for this study).

\subsection{Measurement and Metrics}\label{measurement-and-metrics}

The dependent variable that we need to measure is the energy consumption of the software application, measured in Joules. This gives us a robust indicator and allows us to compare different configurations across varying execution times. Unlike instantaneous power measurements (in Watts), Joules account for the total energy consumed over the course of a full benchmark execution. We also record the execution time of the benchmarks to compare energy consumption with runtime performance. The independent variables in our experiment are Spring Boot and JVM versions. We also controlled for the compilation target version and virtual threads as secondary independent variables in further experiments. All experiments use exactly the same version of Spring Petclinic REST. For the JVM distribution, we selected Eclipse Temurin, a popular, community-managed JDK with frequent releases. See \autoref{tbl:variables} for an overview of all variables. When applying this method, the number of variables and values should be kept limited, as the number of configurations to measure grows quickly due to combinatorial explosion.

\begin{table}
\centering\scriptsize
\begin{tabular}{l l l l}
\toprule
\textbf{Variable} & \textbf{Type} & \textbf{Values} & \textbf{Purpose} \\
\midrule
Spring Boot version & Independent & \{3.0, 3.1, 3.2, 3.3, 3.4\} & Compare framework versions \\
JVM version & Independent & \{17, 21, 23\} & Compare runtime versions \\
Compilation target & Independent & \{21, 23\} & Check compiler influence \\
Virtual threads & Independent & \{enabled, disabled\} & Evaluate scheduling effect \\
Energy (J) & Dependent & \textit{measured} & Main comparison metric \\
Runtime (s) & Dependent & \textit{measured} & Performance comparison \\
\bottomrule
\end{tabular}
\caption{Variables used in the experiments.}
\label{tbl:variables}
\end{table}

We exclude network traffic and disk usage from our measurements, as they are assumed to be identical for all versions being tested. Additionally, we disregard GPU energy usage, as typical Spring applications do not utilize GPUs. Other peripheral devices are also ignored since Spring applications are usually run on servers. This focused approach allows us to isolate the energy impact of the JVM and framework versions specifically.

\subsection{Measurement Procedure Models}\label{measurement-procedure-models}

\begin{figure*}[ht]
\centering
\includegraphics[width=\textwidth]{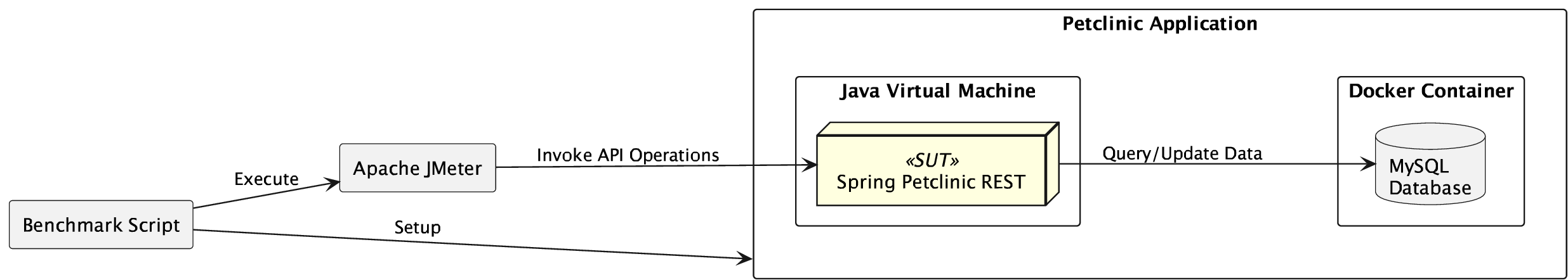}
\caption{A high-level overview of the different benchmark components. The benchmark script sets up the application and starts Apache JMeter, which invokes API operations in the Spring Petclinic REST application. The application interacts with a MySQL database running in a Docker container.}\label{fig:BenchmarkSetup}
\end{figure*}

To obtain a large enough sample size and thus, reliable results, we measured each combination of both JVM and Spring versions repeatedly. This requires a large degree of automation in the measurement procedure. To this end, we use the API endpoints offered by Spring Petclinic REST, mostly in the form of information holder resources and corresponding state retrieval, creation, and transition operations \citep{PatternsForAPIDesign:2022}. Other variants of the sample application exist to show how to use Spring with different frontend technologies. The API variant has the advantage---compared to, for example, an end-used browser application---that we can easily use it with benchmarking tools. Simulating user interactions via API is more robust and easier to reproduce than using a frontend application intended for human users. A commonly used tool for such benchmarks is Apache JMeter, which has an HTTP Request feature that allows us to invoke operations in the API and generate the required load on the application. By default, the Petclinic application database is populated with a small amount of sample data. To make the benchmark more realistic, we created a larger set of fake data to populate the database. For each run, a new MySQL instance is created in a Docker container and populated with this initial dataset, which is the same for all runs. \autoref{fig:BenchmarkSetup} shows the components of the benchmark.

\begin{figure*}[ht]
\centering
\includegraphics[width=0.5\textwidth]{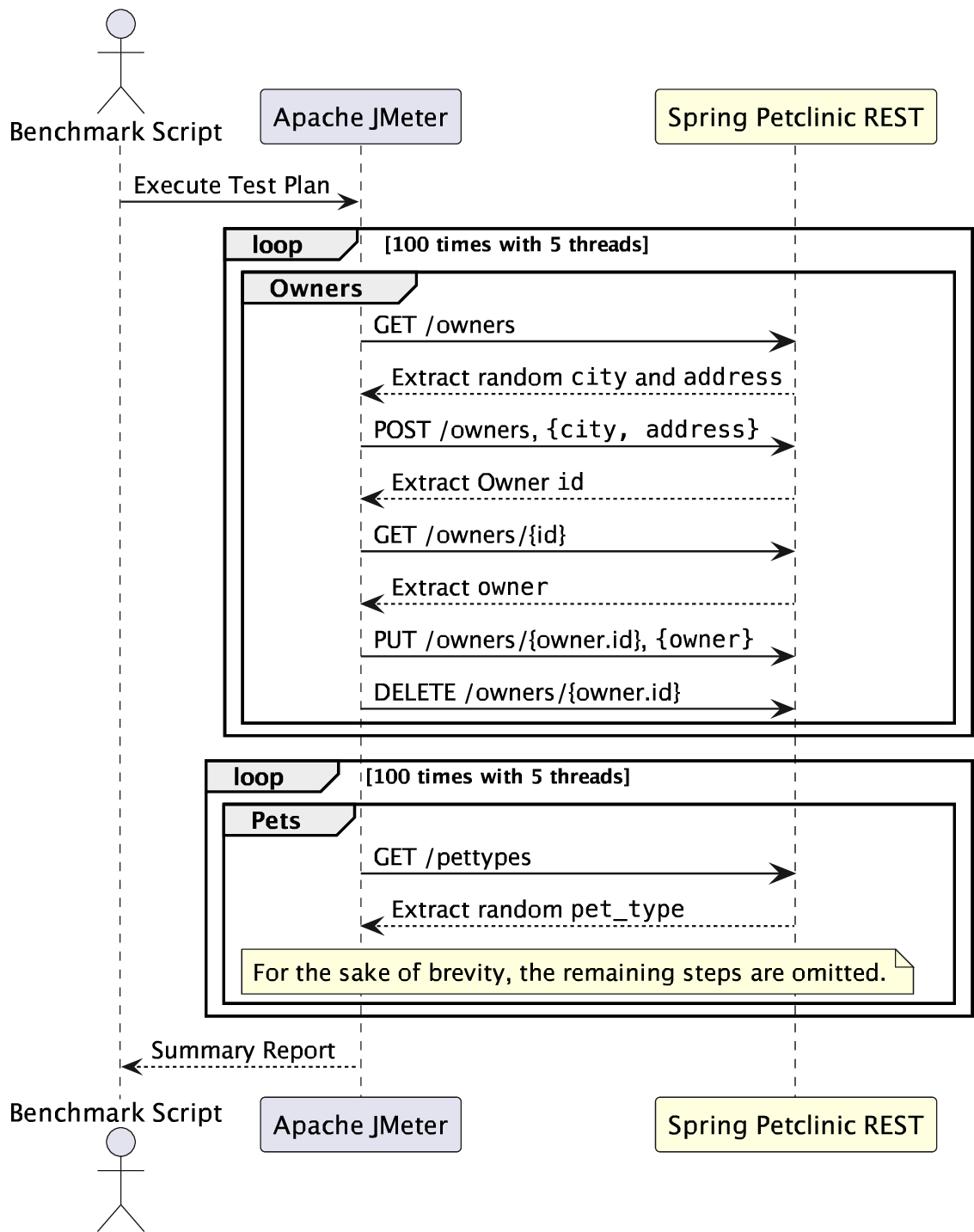}
\caption{Benchmark sequence defined in the JMeter test plan. Each group of operations (Owners, Pets, Visits, Vets) is executed 100 times using 5  threads.}\label{fig:BenchmarkDetails}
\end{figure*}

We developed a benchmark scenario (``test plan'' in JMeter terminology) that uses HTTP GET, POST, PUT, and DELETE requests to interact with the API. The test plan simulates user interactions by creating new owners, adding pets, and updating their information repeatedly. In total, the test plan comprises 5500 GET requests and 2000 POST, 2000 PUT, and 2000 DELETE requests to different operations and resources (Owners, Pets, Visits, etc.) of the API. The high number of requests ensures that the test plan takes several minutes to complete, allowing the JVM to run just-in-time compilation and apply optimizations to the code. See \autoref{fig:BenchmarkDetails} for the sequence of calls.

Spring Boot regularly releases new versions, at the time of writing, there were two major versions (2 and 3) and 6 minor versions (2.7, 3.0, 3.1, 3.2, 3.3, 3.4) available. Supporting version 2 would have required changes to the code, so we restricted the benchmark to version 3 of Spring Boot. The JVM is also regularly updated, with currently two new major releases per year and a long-term supported (LTS) version every two to three years.

For this study, all available LTS Java versions and the latest released version of the Eclipse Temurin distribution were used. For Spring Boot, all 3.x versions that are still supported by the Spring community were used. For each minor version, the latest patch version available at the time was used (e.g.,~3.4.1). Not all Spring Boot releases are compatible with all Java versions, see \autoref{tbl:spring-java-versions} for details.

Newer Java versions contain not only improvements of existing features, but add new features that application developers can take advantage of. These are sometimes purely ``syntactic sugar'' for writing more elegant code (see Project Amber from \citet{ProjectAmber:2025} for several examples) but there are also new libraries that give application and framework developers new possibilities to optimize their software. A recent such addition are virtual threads, ``lightweight threads that dramatically reduce the effort of writing, maintaining, and observing high-throughput concurrent applications,'' introduced in Java 21 \citep{VirtualThreads:2025}. Spring Boot started supporting virtual threads in version 3.4 and can be enabled or disabled (the default) at the start of the application. Since virtual threads suggest efficiency gains especially for blocking, I/O-heavy tasks~\citep{Beronic2021Virtual}, we decided to investigate the effect of this new feature as well.

When compiling Java source code, the compiler generates byte code of a specific target version for the JVM to execute. To run the code, these versions do not have to be identical. For example, byte code compiled with a Java 21 compiler, set to target Java 17, can be run by a Java 23 JVM. When running the benchmark, we always set the compilation target version to the same version of the JVM it will be run with. (We investigate whether the chosen compilation target version is relevant for the energy consumption separately.)

\begin{table}[t]
\centering\scriptsize
\begin{tabular}{@{}lll@{}}
\toprule
Spring Boot Release & Compatible Java Versions & Released     \\ \midrule
3.4.1               & 17, 21, 23               & Dec 19, 2024 \\
3.3.7               & 17, 21                   & Dec 19, 2024 \\
3.2.12              & 17, 21                   & Nov 21, 2024 \\
3.1.12              & 17, 21                   & May 23, 2024 \\
3.0.13              & 17, 21                   & Nov 23, 2023 \\ \bottomrule
\end{tabular}
\caption{Investigated Spring Boot versions and compatible Java versions.}
\label{tbl:spring-java-versions}
\end{table}

\begin{figure*}[ht]
\centering
\includegraphics[width=0.5\textwidth]{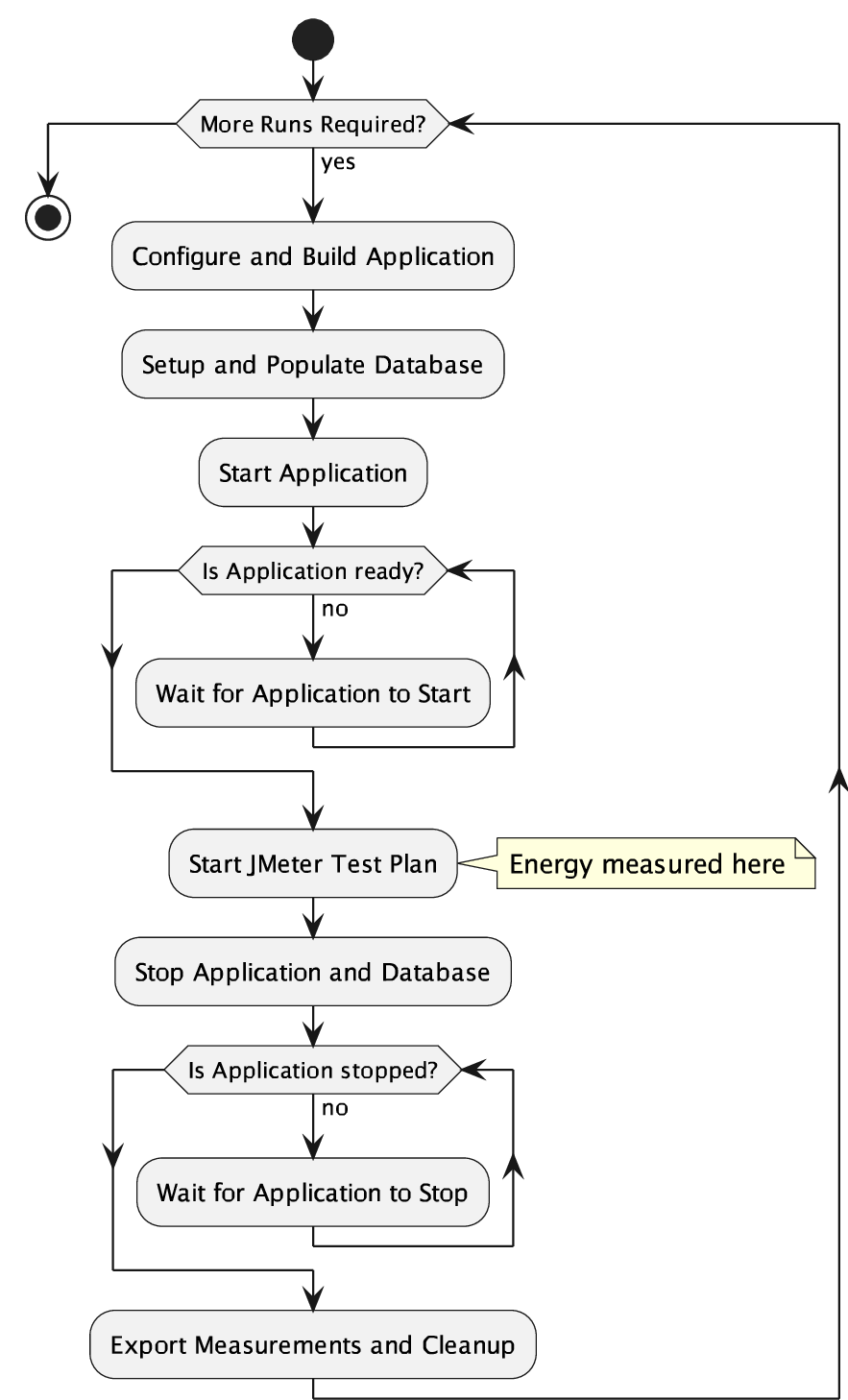}
\caption{Execution steps of a single benchmark configuration run. The process includes building the application, setting up the database, waiting for the application and database to be ready, running the test plan and cleaning up afterward.}\label{fig:BenchmarkSteps}
\end{figure*}

The individual steps of a benchmark run are shown in \autoref{fig:BenchmarkSteps}. For each combination of JVM version and Spring Boot version, this sequence is repeated, and the measurements collected in a CSV file.

A parametrized shell script automates this process and also includes several checks, e.g., if the application is ready and waits accordingly. The scripts, including the setup instructions for Linux and macOS are available on GitHub \citep{SpringPetclinicEnergyTesting:2025}.

\subsection{Measurement Setup}\label{measurement-setup}

\begin{figure}[h!t]
    \centering
    \includegraphics[width=\linewidth]{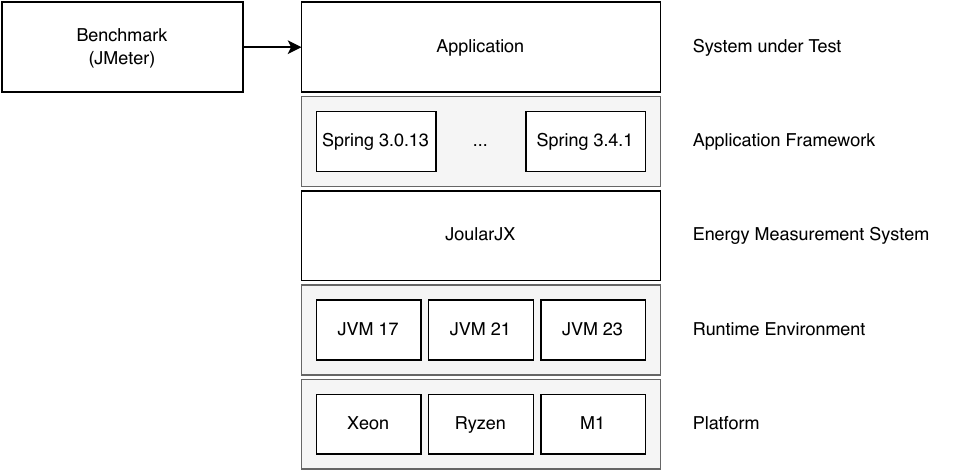}
    \caption{Software and hardware stack for our measurement setup.}
    \label{fig:architecture-java}
\end{figure}

Our concrete measurement setup is visualized as an architecture diagram in \autoref{fig:architecture-java} and explained in the remainder of this section. The energy consumption is measured using the JoularJX tool by \citet{Noureddine:2022}. JoularJX is a Java agent that hooks into the JVM to measure the energy consumption of applications and reports these in Joules. Depending on the platform, Intel's RAPL (Running Average Power Limit) or other hardware counters allow energy measurements through CPU registers. While these only provide the total energy usage, JoularJX then estimates the process energy ($E_{process}$) based on the CPU usage of the process ($CPU_{process}$) in relation to the total CPU usage ($CPU_{total}$) and energy ($E_{system}$) as follows:

\[
E_{process} = \left( \frac{CPU_{process}}{CPU_{total}} \right) \times E_{system}
\]

The project website \citet{JoularJXWebsite:2025} and \citet{Brunnert:2025} provide more details on how JoularJX works. JoularJX supports black-box and white-box measuring with energy usage reported for the whole program or per individual Java method calls. For our purposes, we were only interested in the overall energy usage. The full configuration for JoularJX can be found in the GitHub repository \citep{SpringPetclinicEnergyTesting:2025}.

We decided for such a software-based approach to measuring and not use hardware-based setup---using a power meter---because it allows us to test a wider range of hardware (which we did not have to buy) and also makes the measuring easier by not having to establish and calculate with a baseline of the system under test. A potential disadvantage of the software-based approach is that the  tool can distort the  measurements. However, a study by \citet{Brunnert:2025} found that the overhead of running JoularJX compared to measurements without instrumentation is negligible.

Applications written using Spring Boot are often run in virtual machines on public or private clouds. Unfortunately, the additional abstraction and indirection by the virtual machine makes it harder to collect energy measurements and requires control of the host system that runs the virtual machines, which is not accessible in cloud environments. Instead, we used three different ``bare metal'' cloud servers from Scaleway, a European cloud provider, specifically an \textit{Elastic Metal EM-A115X-SSD} instance with an Intel Xeon CPU E3-1231 v3 @ 3.40GHz CPU and 32 GB RAM (``Xeon'' for short in later figures), an \textit{Elastic Metal EM-A610R-NVME} with an AMD Ryzen PRO 3600 6C/12T 3.6 GHz CPU and 32 GB RAM (``Ryzen''), and a \textit{Bare Metal Apple Server} (Mac mini) with an Apple M1 CPU with 8 cores and 8 GB RAM (``M1'') to include a different CPU architecture as well. This setup makes any hardware variability explicit, allowing us to account for differences in energy consumption across machine types.

\subsection{Data Evaluation Models}\label{data-evaluation-models}

\begin{figure*}[ht]
\centering
\includegraphics[width=1\textwidth]{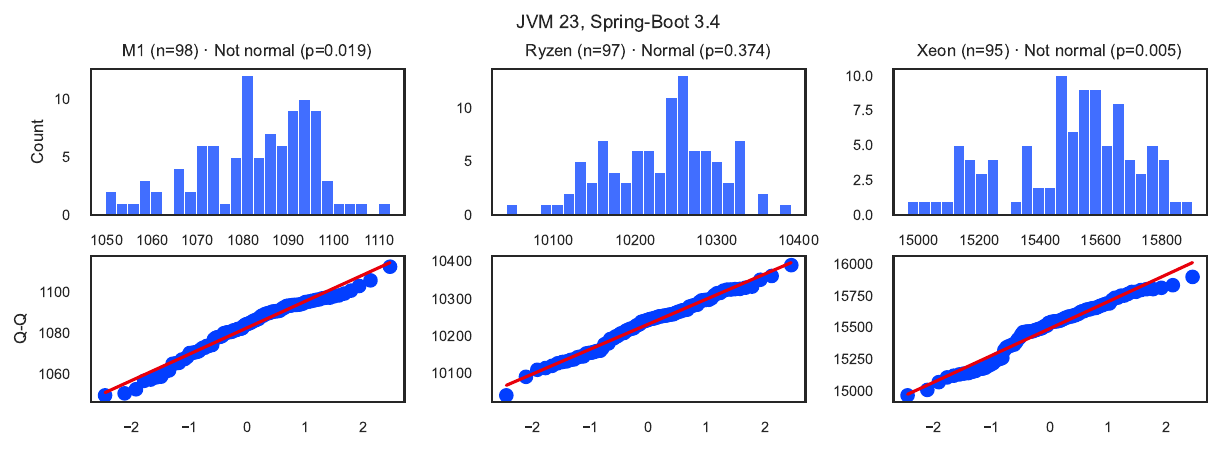}
\caption{Exemplary histograms and Q–Q plots of energy measurements for the M1, Ryzen, and Xeon processors under JVM 23 and Spring Boot 3.4 versions. The top row shows histograms per platform, and the bottom row shows corresponding Q–Q plots. The M1 and Xeon distributions show clear deviations from normality, while Ryzen measurements are approximately normal. These examples illustrate the range of distributional behaviors observed across platforms.}\label{fig:HistogramQQ}
\end{figure*}

After running the experiments doing 100 iterations for each possible combination of Spring Boot and JVM (see \autoref{tbl:number-of-runs}), we collected all the measurements from all systems to start the exploratory data analysis using descriptive statistics and visualizations. To clean the data, we used the interquartile range method (IQR) \citep{Vinutha:2018:IQR}. The IQR finds outliers by measuring the middle half of a dataset, calculated as IQR = Q3 - Q1, where Q1 and Q3 are the first and third quartile. Using the 1.5 × IQR rule, any value below Q1 - 1.5 × IQR or above Q3 + 1.5 × IQR is considered an outlier. After cleaning the data by removing all outliers (for both energy and runtime measurements) from the dataset, between 84 and 100 measurements remained for each combination (see \autoref{tbl:number-of-runs}). We then performed a Shapiro-Wilk test \citep{Shapiro:1965} to check if the data follows a normal distribution. The test compares the data to a perfectly normal distribution and checks how similar they are. It gives a p-value, where a low value means that the data is probably not normal. Interestingly, this showed that while some groups are normally distributed, others show a bimodal distribution with unequal peaks, which we confirmed using histogram and Q-Q plots (see \autoref{fig:HistogramQQ} for a subset of distributions).

We did not further investigate the reasons for the non-normality, but it may reflect inherent variability in the system. Factors might include differences in Linux kernel scheduling, JVM specific aspects such as garbage collection or just-in-time compilation, as well as hardware-specific effects like CPU frequency scaling or thermal throttling. These factors could cause some runs to cluster differently, producing skewed or bimodal distributions.

\begin{table}[t]
\centering\scriptsize
\begin{tabular}{lllllll@{}}
\toprule
             && 3.0.13 & 3.1.12 & 3.2.12 & 3.3.7 & 3.4.1 \\ \midrule
 &17.0.13-tem & 100    & 100    & 100    & 100   & 100   \\
 &21.0.5-tem  & 100    & 100    & 100    & 100   & 100   \\
 &23.0.1-tem  & --     & --     & --     & --    & 100   \\ \midrule
 Xeon&17.0.13-tem & 100    & 100    & 100    & 88& 92\\
 &21.0.5-tem  & 99& 90& 99& 93& 99\\
 &23.0.1-tem  & --     & --     & --     & --    & 95\\ \midrule
 Ryzen&17.0.13-tem & 90    & 97    & 100    & 89 & 92\\
 &21.0.5-tem  & 100& 100& 95& 94& 98\\
 &23.0.1-tem  & --     & --     & --     & --    & 97\\ \midrule
 M1&17.0.13-tem & 87    & 84    & 93    & 90 & 95 \\
 &21.0.5-tem  & 99& 99& 99& 97& 93\\
 &23.0.1-tem  & --     & --     & --      & --     & 98\\ \bottomrule
\end{tabular}
\caption{Overview of benchmark runs for each system and version. We ran the benchmark 100 times on each machine (first row). Note that only
Spring Boot 3.4.1 supports Java 23 as the compilation target. The table also shows the number of remaining measurements for each system after outliers have been removed.}\label{tbl:number-of-runs}
\end{table}

We then used box plots to visually inspect the different groups and a non-parametric Kruskal-Wallis H-test \citep{Kruskal-Wallis:1952}, as the samples are not normally distributed. To learn which groups differ in their medians we also ran Conover's test \citep{Conover:1979} to get pairwise comparisons between groups, helping to identify which specific groups have significant differences in their medians. To measure the effect size of these differences, we calculated Cliff's delta \citep{Cliff:1993} for each pairwise comparison. Cliff's delta quantifies how often values in one group are higher or lower than those in another group, making it well suited for non-parametric data. The scale ranges from -1 to 1, with 0 indicating equality and 1 that the first group dominates the second. We use heatmaps to visualize these results.

All the data, including raw measurements, and analysis scripts that produce the figures in this paper are available on Zenodo \citep{ZenodoDataset:2025}.

\section{Results}\label{results}


In this section, we will answer the research question using the results from our case study. Through the substitution of different component versions and the benchmarking of these variations, we gained insight into how the different versions behave.

\subsection{Middleware Application Framework Energy Consumption}\label{spring-framework-energy-consumption-rq1}

\begin{figure*}[ht!]
\centering
\includegraphics[width=\textwidth]{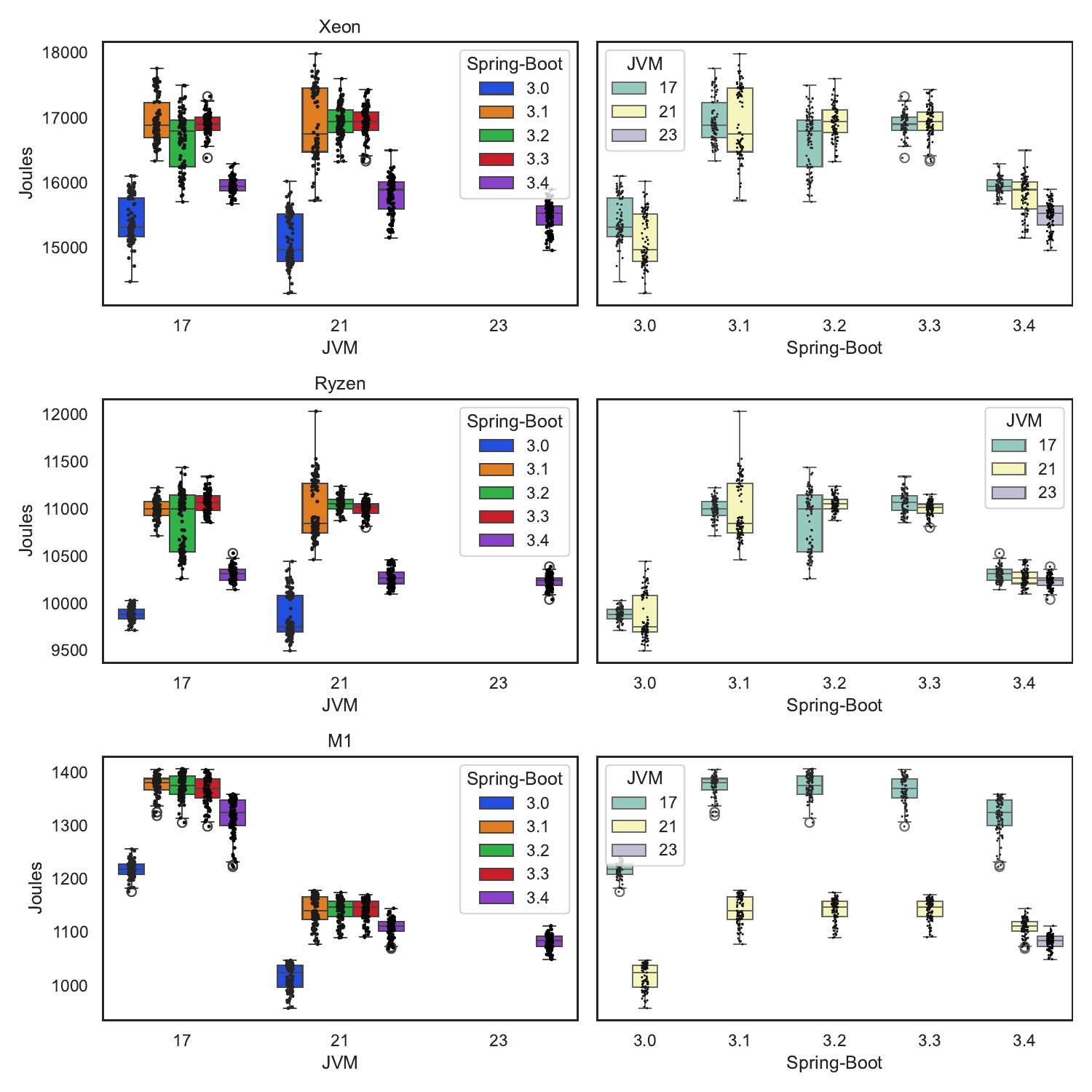}
\caption{Energy consumption across JVM and Spring Boot versions on three systems (Xeon, Ryzen, M1). Left column: grouped by JVM version and colored by Spring Boot version. Right column: grouped by Spring Boot version and colored by JVM version. Each box plot shows the distribution of energy usage (in Joules) across benchmark runs.}\label{fig:joules-by-spring-boot-and-jvm-versions}
\end{figure*}

Our first experiment concerns the evolution of energy consumption across different versions of the middleware application framework, in the hope that newer versions are getting more efficient. The results for the framework chosen for our study (Spring Boot) are shown in the left column of \autoref{fig:joules-by-spring-boot-and-jvm-versions}. We can see that this is not the case: the oldest version, Spring Boot 3.0.13, is using the least amount of energy on all three systems and the 17 and 21 JVM versions (being a somewhat older version, it does not support Java 23 as a compilation target). Versions 3.1 to 3.3 perform similarly, whereas the latest version, 3.4, is more efficient again, but still not as good as 3.0.

\begin{table}
\centering\scriptsize

\begin{tabular}{llllllllll}
\toprule
& & \multicolumn{2}{l}{Xeon}                & \multicolumn{2}{l}{Ryzen}               & \multicolumn{2}{l}{M1}                \\ \midrule
\multicolumn{2}{l}{Spring Boot}                              & JVM 17             & JVM 21             & JVM 17             & JVM 21             & JVM 17            & JVM 21            \\ \midrule
3.0            & 3.1            & \textbf{4.21e-103} & \textbf{1.19e-97}  & \textbf{5.20e-100} & \textbf{6.33e-101} & \textbf{3.99e-84} & \textbf{3.33e-80} \\
3.0            & 3.2            & \textbf{3.96e-79}  & \textbf{3.02e-103} & \textbf{1.70e-97}  & \textbf{1.74e-124} & \textbf{4.99e-86} & \textbf{4.46e-77} \\
3.0            & 3.3            & \textbf{2.87e-99}  & \textbf{7.94e-103} & \textbf{2.65e-117} & \textbf{1.10e-108} & \textbf{2.31e-75} & \textbf{3.40e-78} \\
3.0            & 3.4            & \textbf{2.20e-14}  & \textbf{2.55e-15}  & \textbf{3.17e-19}  & \textbf{1.33e-17}  & \textbf{3.30e-21} & \textbf{2.23e-24} \\
3.1            & 3.2            & \textbf{1.11e-06}  & 1.00e+00           & 4.34e-01           & \textbf{1.21e-07}  & 8.44e-01          & 1.00e+00          \\
3.1            & 3.3            & 9.40e-01           & 1.00e+00           & \textbf{1.24e-05}  & \textbf{3.40e-02}  & 5.86e-02          & 1.00e+00          \\
3.1            & 3.4            & \textbf{1.60e-62}  & \textbf{2.31e-59}  & \textbf{8.15e-56}  & \textbf{1.57e-57}  & \textbf{3.24e-41} & \textbf{1.43e-28} \\
3.2            & 3.3            & \textbf{1.18e-06}  & 1.00e+00           & \textbf{3.71e-07}  & \textbf{1.49e-03}  & 5.86e-02          & 1.00e+00          \\
3.2            & 3.4            & \textbf{1.79e-39}  & \textbf{3.66e-64}  & \textbf{5.96e-53}  & \textbf{1.08e-82}  & \textbf{5.02e-42} & \textbf{4.58e-26} \\
3.3            & 3.4            & \textbf{4.72e-60}  & \textbf{3.07e-64}  & \textbf{5.19e-75}  & \textbf{3.96e-66}  & \textbf{3.30e-32} & \textbf{3.61e-27} \\ \bottomrule
\end{tabular}

\caption{Conover Post-hoc Test Results (Holm-adjusted p-values) comparing Spring Boot versions (first column) on Xeon, Ryzen and M1 with JVM 17 and 21, respectively. Bold values indicate statistically significant differences (p < 0.05).}
\label{tbl:effect-size-by-spring-boot-version-conover-results}
\end{table}

\begin{figure*}[ht!]
\centering
\includegraphics[width=0.75\textwidth]{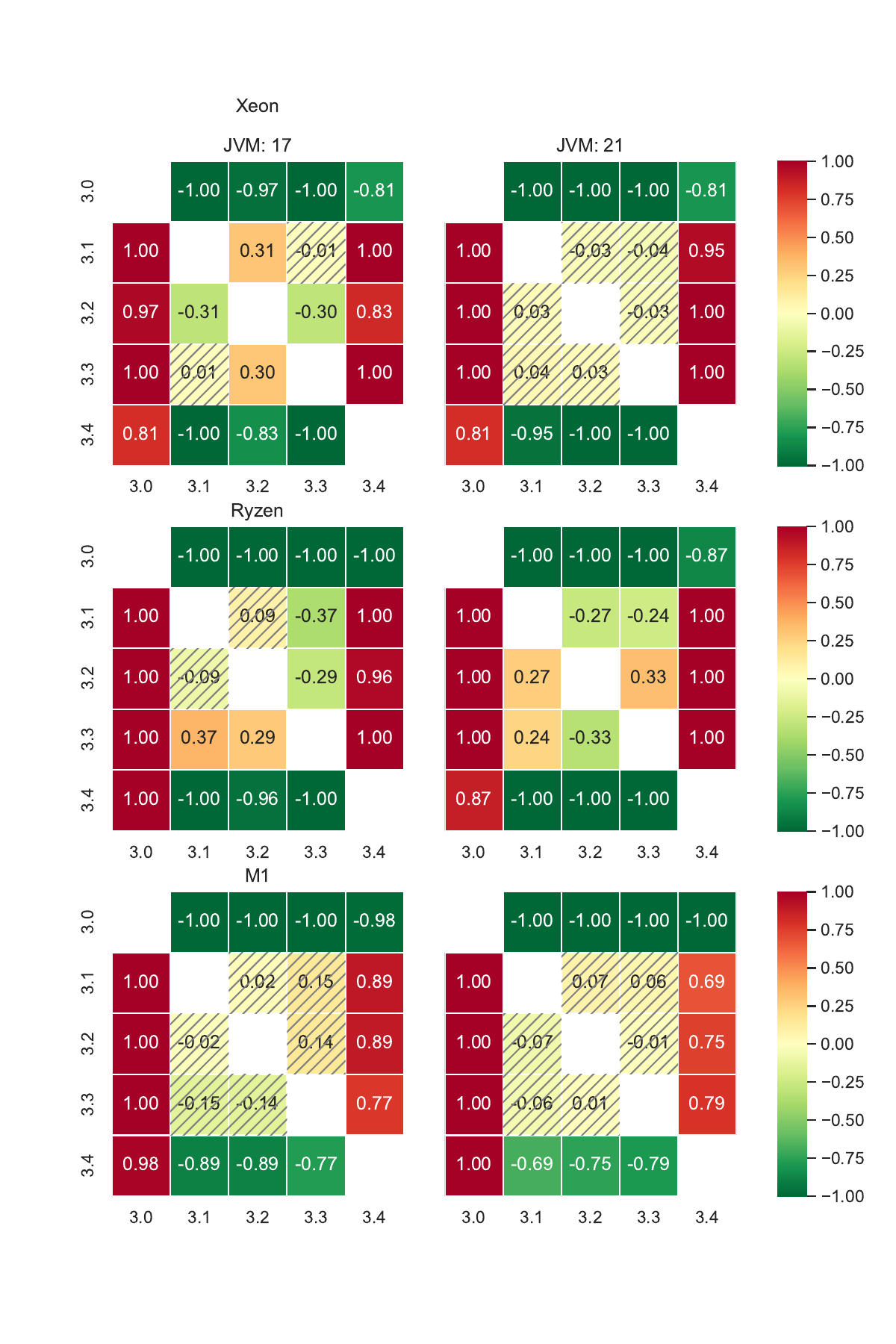}
\caption{Effect sizes measured using Cliff's Delta for pairwise comparisons of Spring Boot versions on three systems (Xeon, Ryzen and M1) and two JVM versions (17 and 21). Each cell in the heatmap quantifies the difference in energy consumption between the row and column versions. Positive (red) values indicate higher energy usage for the row version, while negative (green) values indicate the opposite. Values close to 0 (yellow) suggest little difference between versions. Shaded results indicate non-significant differences (p > 0.05).}\label{fig:effect-size-by-spring-boot-version}
\end{figure*}

To quantify these differences, we ran Conover's test (see \autoref{tbl:effect-size-by-spring-boot-version-conover-results} for details) and calculated Cliff's delta and plotted this data using heatmaps, shown in \autoref{fig:effect-size-by-spring-boot-version}. For each of the three systems, we show one heatmap per JVM. In each heatmap, the cells show the comparison between two Spring Boot versions. A positive value indicates that energy consumption of the version of the row is larger than the version of the column. For example, if we take the lower left value of 0.98. This means that on the M1 system, using a JVM 17, the Spring Boot version 3.4.1 consumes significantly more energy than version 3.0.13. If we move one cell to the right, the -0.89 shows that 3.4.1 is more efficient than version 3.1.12. The diagonal is always empty since these values are 0. Note that the grid is also skew-symmetric. The box plots showed that versions 3.1 to 3.3 are similar to each other, this can also be seen in the center of the heatmaps, where the values are closer to 0.

\subsection{Energy Consumption of the Runtime Environment}\label{jvm-energy-consumption-rq2}

Switching perspectives, we now compare the effect of the JVM while keeping the Spring Boot version the same. This is shown in the second column of \autoref{fig:joules-by-spring-boot-and-jvm-versions}. Comparing the box plots for each Spring Boot version and measured system, we can see that there seems to be a tendency for newer JVMs to use less energy, but the results are quite different between the systems that ran the benchmarks. On the Xeon and Ryzen systems, there is no significant difference between the JVMs. Using Spring Boot 3.4.1, we have three JVMs to compare and here we see a tendency for newer versions to consume less energy. On M1, the differences between JVMs are striking, with newer versions clearly using less energy. 

\begin{figure}[ht]
\centering
\includegraphics[width=0.7\textwidth]{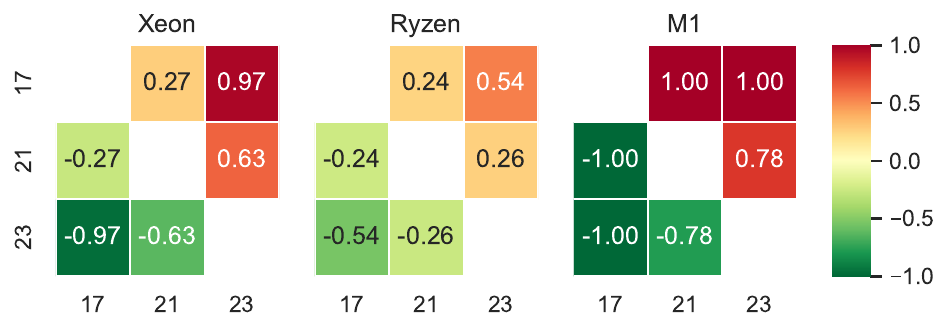}
\caption{Effect size measured using Cliff's Delta for Spring Boot 3.4. Positive values indicate that the row JVM consumes more energy than the column JVM.}\label{fig:effect-size-by-jvm-version}
\end{figure}

We again ran Conover's test (see \autoref{tbl:effect-size-by-jvm-version-conover-results} for details), calculated Cliff's delta, and plotted this data using heatmaps, shown in \autoref{fig:effect-size-by-jvm-version}. Results are only shown for Spring Boot 3.4, where there are three JVM versions to compare. 


\begin{table}[ht]
\centering\scriptsize
\begin{tabular}{lllll}
\toprule
\multicolumn{2}{l}{JVM} & Xeon Spring-Boot 3.4 & Ryzen Spring-Boot 3.4 & M1 Spring-Boot 3.4 \\
\midrule
17             & 21             & \textbf{3.77e-06}    & \textbf{1.15e-03}     & \textbf{1.38e-53}  \\
17             & 23             & \textbf{8.85e-38}    & \textbf{2.35e-10}     & \textbf{3.80e-100} \\
21             & 23             & \textbf{3.66e-22}    & \textbf{9.39e-04}     & \textbf{1.45e-33}  \\
\bottomrule
\end{tabular}
\caption{Conover Post-hoc Test Results. Bold values indicate statistically significant differences (p < 0.05).}
\label{tbl:effect-size-by-jvm-version-conover-results}
\end{table}

\subsection{Compilation Target}\label{compilation-target-rq2b}

\begin{figure}[ht]
\centering
\includegraphics[width=\textwidth]{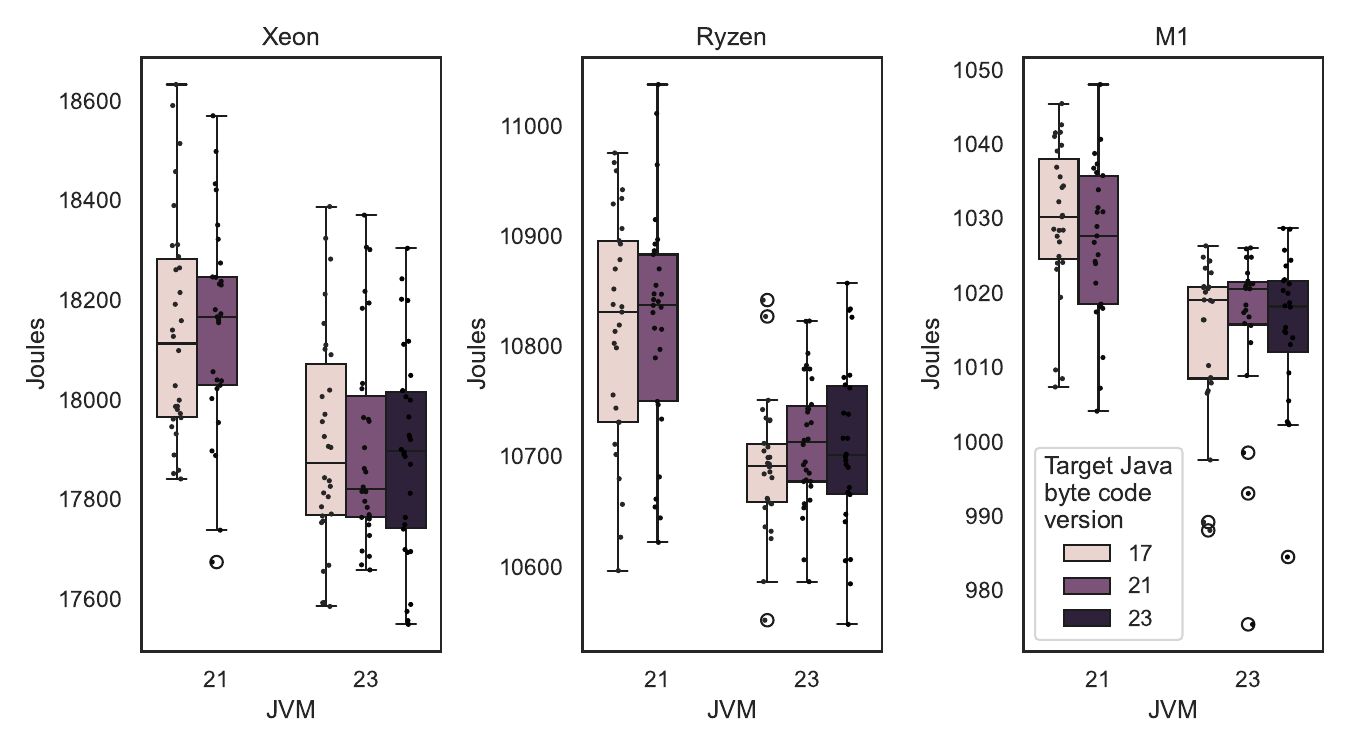}
\caption{Box plots comparing the energy consumption when compiling to different Java byte code versions running on a JVM 21/23 on the three systems, indicating no significant difference in energy usage between versions.}\label{fig:varying-compilation-target}
\end{figure}

One assumption we made when setting up the benchmark was that the targeted compilation version of Java does not affect the energy consumption and that the difference is only due to the JVM version running the byte code. To verify these assumptions, we performed a variation of the benchmark using the Xeon, Ryzen and M1 systems, Spring Boot 3.4 and the JVM in version 21 and 23, but with the compilation target set to Java 17, 21, and 23, respectively. The energy consumption distributions, shown in \autoref{fig:varying-compilation-target}, are very similar, with overlapping medians and interquartile ranges, indicating no significant difference in energy usage between versions. We confirm this by a Kruskal-Wallis test, the results are shown in \autoref{tbl:varying-compilation-target}.

\begin{table}[ht]
\centering\scriptsize
\begin{tabular}{llrllrr}
\toprule
Machine & JVM & N & Group sizes by Java version & Median energy in Joules & H & p-value \\
\midrule
Xeon & 21 & 59 & 17: 30, 21: 29 & 17: 18113, 21: 18167 & 0.486 & 0.486 \\
Xeon & 23 & 90 & 17: 30, 21: 30, 23: 30 & 17: 17874, 21: 17821, 23: 17898 & 0.090 & 0.956 \\
Ryzen & 21 & 58 & 17: 29, 21: 29 & 17: 10831, 21: 10837 & 0.003 & 0.957 \\
Ryzen & 23 & 86 & 17: 30, 21: 30, 23: 26 & 17: 10691, 21: 10713, 23: 10701 & 2.193 & 0.334 \\
M1 & 21 & 52 & 17: 27, 21: 25 & 17: 1030, 21: 1028 & 1.228 & 0.268 \\
M1 & 23 & 72 & 17: 25, 21: 23, 23: 24 & 17: 1019, 21: 1021, 23: 1018 & 0.542 & 0.763 \\
\bottomrule
\end{tabular}
\caption{Kruskal-Wallis test results by machine and JVM. It shows that the compilation target (Java 17, 21, and 23) does not significantly impact the energy consumption.}
\label{tbl:varying-compilation-target}
\end{table}

\subsection{Execution Paradigms/Semantics}\label{virtual-threads-rq2c}

\begin{figure}[ht]
\centering
\includegraphics[width=\textwidth]{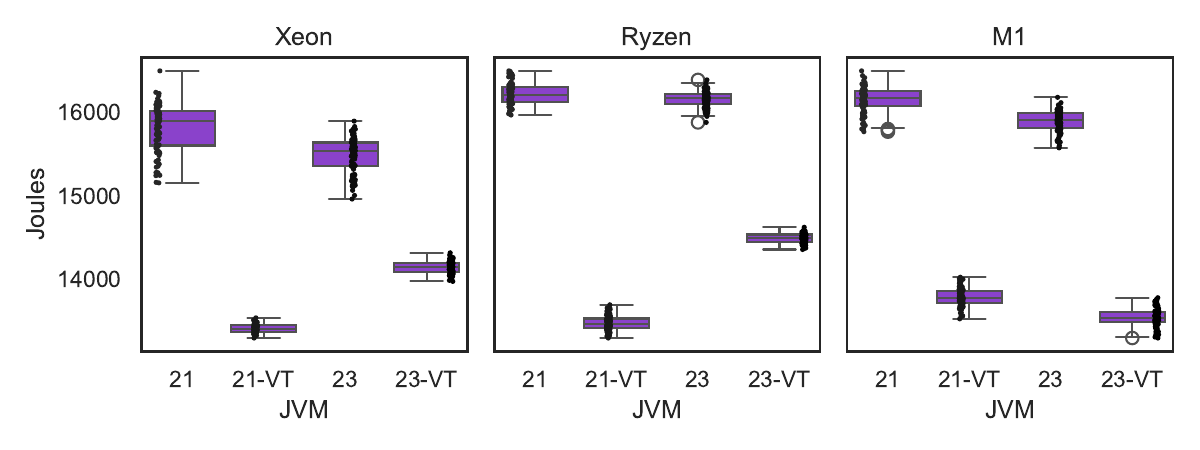}
\caption{Energy consumption with and without virtual threads on three systems (Xeon, Ryzen, M1), using Spring Boot 3.4. The ``-VT'' labels indicate benchmarks executed with virtual threads enabled. Both Java 21 and 23 show substantial energy savings when virtual threads are used.}\label{fig:joules-by-spring-boot-and-jvm-versions-virtual-threads}
\end{figure}

To find out if changes of execution semantics have an impact on energy consumption, we investigated the new Java feature of virtual threads \citep{VirtualThreads:2025}. Virtual threads (VT) are lightweight, user-mode (i.e., not managed by the operating system but by the JVM) threads in Java that allow concurrent tasks to run efficiently with minimal resource overhead. Spring supports virtual threads since version 3.4, but they are disabled by default and need to be enabled by the developer. No other changes to the code are necessary. To investigate this effect, we ran the benchmark again with JVM version 21 and 23 (17 does not support virtual threads) and Spring Boot 3.4. The results are shown in \autoref{fig:joules-by-spring-boot-and-jvm-versions-virtual-threads}. In all cases, the VT-enabled versions (21-VT, 23-VT) consume significantly less energy than their non-VT counterparts.

One notable difference between 21-VT and 23-VT on Xeon and Ryzen compared to M1 is that the latest version only shows improved performance on M1. One possible explanation is that M1, as a relatively new architecture, may still be receiving Java optimizations specific to Apple's hardware. In addition, M1's aggressive power management may improve the efficiency of the updated JVM threading model more effectively than on traditional processors.

\section{Discussion}\label{discussion}

In this section, we discuss the implications of our results for energy efficiency and decision-making for sustainable software engineering.

\subsection{Interpretation of the Results}
Our results show that newer Spring Boot versions do not consistently improve energy efficiency. The oldest version, 3.0, consumes the least energy of all tested systems and JVM versions. Versions 3.1 through 3.3 perform similarly, while 3.4 shows some improvement, but still does not match the efficiency of 3.0. This suggests that changes introduced in later Spring Boot versions may have increased energy consumption, possibly due to added features or internal changes that affect resource usage. Further investigation is needed to identify specific changes in Spring Boot that affect energy efficiency, but this is out of scope of this paper.

Comparing JVM versions while keeping the Spring Boot version constant shows that newer JVMs tend to consume less power, but the results vary from system to system. On Xeon and Ryzen, the differences in power consumption between JVMs are relatively small, with newer versions showing improvements. On M1, the reduction in energy consumption with newer JVMs is much more pronounced. JVM performance optimizations, such as reductions in garbage collection pause times introduced in OpenJDK 22 \citep{JEP423}, could have lowered CPU utilization and thus energy consumption. Since the M1 is a relatively new platform, it is plausible that optimizations for this architecture yield noticeable gains. Additionally, because the M1 features an integrated CPU with performance and efficiency cores, memory, and power‑management subsystem, such improvements may result in lower energy use. 

Comparing the energy consumption of the M1 system versus Xeon/Ryzen (see \autoref{fig:joules-by-spring-boot-and-jvm-versions}), we see an apparent order of magnitude difference. One plausible explanation for this difference lies in the fundamentally different design goals of the underlying CPU architectures. Xeon and Ryzen processors are engineered primarily for high absolute performance and scalability. In contrast, M1 is optimized for performance per Watt as a mobile-class system on a chip designed for efficiency through tight integration and a heterogeneous core design similar to the Arm \emph{big.LITTLE} architecture. \citet{Graham:2022} developed an AArch64 JVM for Arm Cortex-A devices and have shown that ``in most cases the A7x cores display comparable, or better performance than their x86-64 counterpart.'' While Cortex-A and M1 use different microarchitectures, they do share some architectural characteristics, so results observed on Cortex-A systems might also help explain the behavior seen on M1. 

However, another and likely more important factor that explains the difference is that the energy‑measurement mechanisms used by JoularJX  differ substantially across systems. Xeon and Ryzen use RAPL, whereas M1 relies on macOS Powermetrics, which are not comparable as they measure different system boundaries. So the apparent differences may not reflect real hardware‑efficiency differences, but also differences in measurement scope. Whether this is a real difference in energy efficiency or not is still an open question, which we discuss in \autoref{future-work}. 

We have also shown that the compilation target does not have a significant impact. This is useful to know because it means that developers only need to upgrade their production JVMs without having to change their build scripts and development systems. In addition, virtual threads should be enabled for CRUD-focused applications such as PetClinic, even if it means upgrading to a newer version of Spring Boot, which will likely include code changes.

\begin{figure}[ht]
\centering
\includegraphics[width=\textwidth]{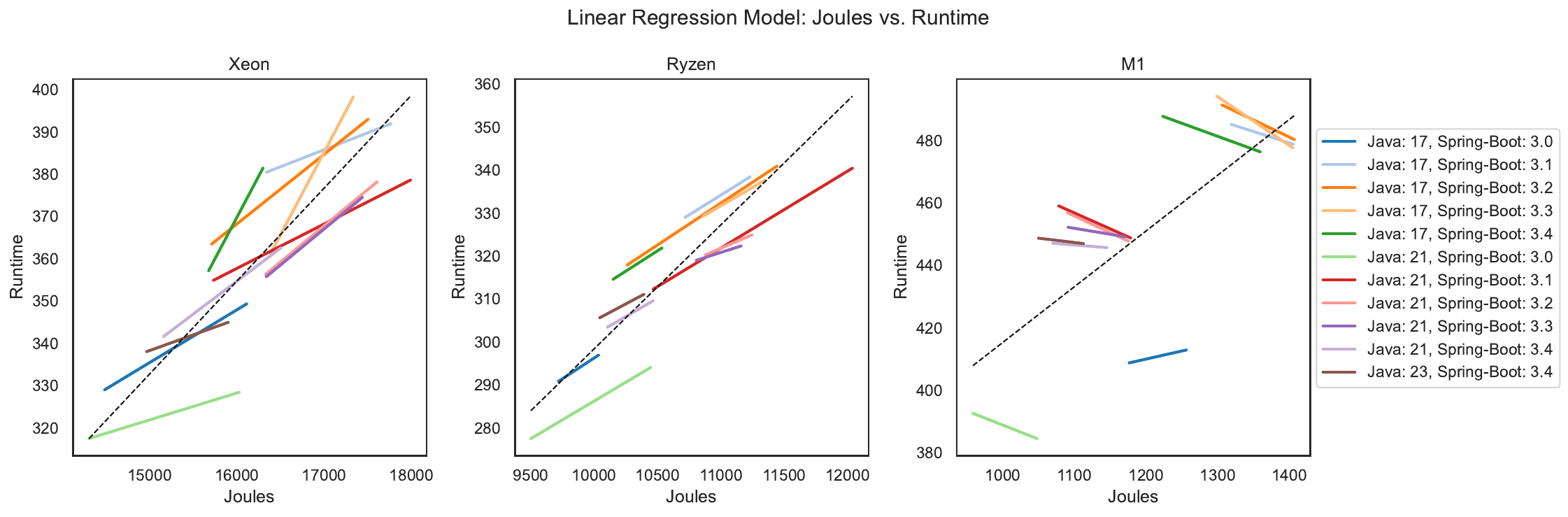}
\caption{Runtime and energy correlation on the different measured systems. Xeon and Ryzen show a clear correlation, whereas M1 exhibits more complex results. On aggregate, it also shows that there is an overall correlation (dashed line) between runtime and energy usage, but many individual correlations show a reversed relation.}\label{fig:runtime-joules-correlations}
\end{figure}

We further analyzed the correlation between the runtime and energy usage of the benchmarks by computing Pearson correlation coefficients for each platform, shown in \autoref{tbl:runtime-joules-correlations} (see \autoref{fig:runtime-joules-correlations} for a graphical analysis). The aggregate platform correlations are strong and positive. At the per-configuration level, Ryzen and Xeon largely retain strong positive correlations across nearly all Java and Spring Boot combinations. Interestingly, M1 shows a different pattern. While the platform‑level correlation is positive, many per‑configuration correlations are negative (often moderate in magnitude).

For the Ryzen and Xeon systems, the correlations indicate the expected relationship: runs that take longer also consume more energy. This would likely change with a different benchmark, where there is a lot of waiting involved, which increases runtime but not necessarily the energy usage. The M1 results are more complex. While the aggregated data show a positive overall correlation, in many individual configurations the relationship reverses. This might be caused by the larger number of CPU cores in the M1 system, different CPU frequency scaling, or how work is scheduled on the M1's Apple Silicon performance and efficiency cores. Relying only on platform‑level summaries can therefore be misleading because the aggregation can mask opposing trends within subgroups.

\begin{table}[ht]
\centering\scriptsize
\begin{tabular}{lccc}
\toprule
 & Xeon & Ryzen & M1 \\
\midrule
All & \textbf{0.809} & \textbf{0.904} & \textbf{0.755} \\
Java: 17, Spring-Boot: 3.0 & \textbf{0.448} & \textbf{0.722} & \textbf{0.518} \\
Java: 17, Spring-Boot: 3.1 & \textbf{0.296} & \textbf{0.861} & \textbf{-0.544} \\
Java: 17, Spring-Boot: 3.2 & \textbf{0.629} & \textbf{0.986} & \textbf{-0.397} \\
Java: 17, Spring-Boot: 3.3 & \textbf{0.641} & \textbf{0.861} & \textbf{-0.671} \\
Java: 17, Spring-Boot: 3.4 & \textbf{0.509} & \textbf{0.748} & \textbf{-0.464} \\
Java: 21, Spring-Boot: 3.0 & \textbf{0.327} & \textbf{0.941} & \textbf{-0.577} \\
Java: 21, Spring-Boot: 3.1 & \textbf{0.546} & \textbf{0.967} & \textbf{-0.594} \\
Java: 21, Spring-Boot: 3.2 & \textbf{0.473} & \textbf{0.651} & \textbf{-0.489} \\
Java: 21, Spring-Boot: 3.3 & \textbf{0.435} & \textbf{0.524} & \textbf{-0.223} \\
Java: 21, Spring-Boot: 3.4 & \textbf{0.493} & \textbf{0.753} & -0.081 \\
Java: 23, Spring-Boot: 3.4 & \textbf{0.847} & \textbf{0.573} & -0.143 \\
\bottomrule
\end{tabular}
\caption{Pearson correlation (Runtime vs Joules). Bold values indicate statistically significant correlations (p < 0.05).}
\label{tbl:runtime-joules-correlations}
\end{table}

\subsection{Implications for Software Stack Upgrades}
Taking informed decisions when choosing the right versions of the JVM and Spring Boot matters when considering the overall energy consumption of the system.
For example, given that the average runtime of the benchmark being 6.3 minutes (Spring Boot 3.1 on Xeon) per 11'500 API calls (one execution of our test), we can extrapolate that 229 of such API call sequences can be run per day. 
As shown, a single execution of our tests on Spring Boot 3.1 can consume up to 17'000\,J on typical server hardware (Xeon). 
Thus, the energy consumption of one server continuously handling API calls as in our tests can consume up to 1'080\,Wh per day, or 394\,kWh per year. 
Assuming a carbon intensity of 300\,g\cotwo/kWh, this results in 118\,kg of \cotwo emissions per year. 
Upgrading to Spring Boot 3.4 (consuming roughly 15'750\,J per test run) would save around 8\,kg of \cotwo emissions per year.
Enabling virtual threads would save another 13\,kg of \cotwo emissions per year.

Thus, upgrading these components can make a significant impact on the energy consumption of the overall software. 
However, optimizing the energy consumption must  be considered in a wider software engineering context, where it is an \emph{additional} criterion for making upgrade decisions: 
upgrading to a more energy-efficient version of the JVM or Spring Boot may come with additional effort for adapting the higher layers of the software stack (e.g., for replacing the use of deprecated API) and testing these adaptations, which consumes energy itself. 
Also, newer versions of software components may introduce new bugs or security vulnerabilities that affect the functionality of the software system.

For large software systems that scale for thousands or even millions of simultaneous users (and thus, thousands of server instances), reducing the energy consumption should be an important factor when making upgrade decisions. 
There are many examples of such systems such as scientific simulation, social media, large language models, or video conferencing. 
Our method of measuring the energy consumption of individual components of the software stack allows organizations to do trade-off analysis, enhances transparency, and supports informed decisions in which energy consumption can be quantitatively considered for maintenance and evolution tasks.

\section{Threats to Validity}\label{threats-to-validity}

In this section, we discuss potential threats to the validity of our study according to \citet{Wohlin:2012}, differentiating between construct, internal, and external validity.

Our research question is broader than the scope of the study. While our study shows that the energy consumption of different versions (and combinations of) application frameworks, runtime environment, and execution pragmatics can change in unexpected ways for Java-based applications, it does not fully address all aspects implied by the research question (e.g., for other kinds of frameworks, or even any kinds of frameworks, programming languages or runtimes). This limits the generalizability of our conclusions and represents a construct validity threat. However, our approach is sufficiently generic such that it can be applied to other kinds of ecosystems and frameworks (e.g., .NET) with minor adaptations.

Concerning internal and construct validity, to mitigate threats to the reliability of the measurements, we performed the benchmarks on three different systems, which run only a base Ubuntu or macOS system with no other user applications running concurrently. Using the high number of repetitions of the whole benchmark and also repetitions within each test plan (in total, the benchmarks took about five days per system to complete). In combination with removing outliers, as described above, fluctuations caused by the underlying system, as well as JVM warm-up effects and garbage collection, should be eliminated. Using cloud servers prevented us from measuring the energy consumption using external power meters, instead, we used JoularJX, which allows real-time measurements, but does not include all the system's components, e.g., the disk or network. 
We also cannot compare different systems to each other, which is not a problem since we are only interested in relative changes within each system as we switched versions. Since the same amount of data is exchanged in all benchmarked versions, there should be no difference caused by disk access and network transfer. However, switching versions could still indirectly affect components not measured, such as disk or network usage, although this is unlikely for minor version changes. Both Spring and the JVM are known for strong backwards compatibility.

While we did vary the JVM versions, we did not investigate if default settings of the JVM changed with newer versions. For example, changes to the default garbage collector might have changed our results. We did verify that the JVM used the same memory settings on each machine. Future work could improve internal validity by systematically evaluating JVM settings, for example, regarding garbage collection details, and their impact on energy efficiency. We also focused solely on the Eclipse Temurin distribution of the JVM. Running the benchmarks with other JVM versions and distributions may produce different results.

Benchmark specifics, such as the specific calls and operations invoked and the chosen system under test, might affect the external validity of our results. If the chosen workload and operations do not reflect real-world usage patterns, the results may not be representative of actual application performance. By repeating the experiments using different JMeter test plans, the results could be validated against other workloads and usage patterns. 

The Spring Petclinic application is a widely used sample application and has been used by other studies on energy consumption and performance \citep{Zhao:2025}, but it may not cover all aspects of real-world Spring applications. Future work could include more complex applications or different domains to improve external validity.

\section{Conclusion}\label{conclusion}

This paper presented a systematic approach for measuring how different versions of software stack components---specifically Spring Boot, the Java Virtual Machine (JVM), and hardware/operating system platforms---influence the energy consumption of a representative application. Our goal was to assess whether automated measurements and controlled component substitution can reveal meaningful and actionable differences in energy usage across software stack versions.

Our empirical results confirm that such differences are both significant and sometimes counterintuitive. Energy consumption varied noticeably across Spring Boot and JVM versions, as well as across hardware architectures. Newer JVM releases consistently reduced energy usage on all platforms considered, with the largest improvements observed on the Apple~M1 processor. Moreover, enabling virtual threads (available from Java~21 onward) yielded substantial energy savings without requiring any changes to the application code. In contrast, the compilation target version had no measurable effect on energy consumption, simplifying upgrade decisions for developers and organizations.

These findings demonstrate that energy consumption can serve as a practical and evidence-based decision criterion in software maintenance and evolution. By upgrading selectively to more energy-efficient runtime components, organizations can improve the sustainability of their systems without modifying application logic or architecture.

While our study focused on a single application and a selected set of platforms, the proposed measurement approach is broadly reusable and can be applied to other applications, frameworks, and hardware configurations. Our methodology and tooling support the integration of energy measurements into existing benchmarking and evaluation pipelines, enabling more informed and sustainable upgrade decisions.

In summary, energy-aware decision making is both feasible and valuable. Incorporating energy measurements into software evolution processes can help organizations make upgrades that not only improve performance and maintainability, but also reduce the environmental footprint of their systems.

\section{Future Work}\label{future-work}

This section outlines our current and planned research directions to deepen the understanding of how software design decisions and system components affect energy consumption.

\subsection{Software Architecture and Energy Consumption}
One area of research we are pursuing is the impact of different API patterns and API refactoring on energy consumption. For example, over-fetching large response payloads, deeply nested structures, or redundant requests due to missing support for request bundles can lead to increased CPU usage, memory pressure and network traffic. These effects can build up, particularly in high-throughput backend systems. The Introduce Pagination refactoring \citep{Stocker:2023}, for example, can reduce the volume of data transferred and processed per request, potentially lowering energy use. Similarly, consolidating multiple fine-grained endpoints into coarse-grained operations can reduce the number of network round trips and improve energy efficiency. This will require a more complex setup that also includes a measurement or approximation for network efficiency \citep{Bister:2024}.

The choice of response format could also play a role, as the costs of serialization and parsing differ between formats or implementations thereof. While such design choices are usually made for reasons of performance or maintainability, their energy implications remain under-explored. Future work could involve isolating and benchmarking common API design patterns in order to quantify their contribution to energy consumption.

We also plan to investigate architectural refactoring towards serverless cloud systems, where code runs for short periods and resources are metered. Changes that cut computation, memory usage, or data transfer can lead to lower energy and also cost. Some abstractions or patterns like unnecessary layering, repeated data transformation, or excessive object mapping introduce overhead. Future work should identify such energy-related antipatterns, which refactorings help and which hurt, and provide practical guidance for energy-aware code changes.

\subsection{Hardware-specific Measurements}

We have seen significant differences of the energy consumption of the same application on different hardware platforms.
In future research, we plan to compare the absolute power consumption of the systems, measured using an external power meter to determine whether the M1 is really that much more efficient or if the difference is caused by the different underlying measurement methods (Intel/Xeon RAPL vs macOS Powermetrics).

\subsection{Decision-making Frameworks}

We want to integrate our energy measurement method into existing frameworks for decision-making in software development such as \cite{krishnan_decision_2004} or \cite{wohlin_towards_2021}.
This will allow practitioners to document the energy measurements and improve transparency when making upgrade decisions.
Also, combining our approach with the SEEDS framework (\cite{manotas_seeds_2014}) would allow developers to optimize their code itself with respect to energy consumption (e.g., replacing an ArrayList with a LinkedList for a given use case) and test if the chosen optimization persists across different versions of the underlying stack.

\subsection{Underlying Causes of Energy Differences}

For most practical upgrade scenarios, it should be sufficient to know to what extent different versions of the software stack (JVM, Spring) differ in terms of the energy consumption of the whole application. Investigating the underlying causes, however, could provide valuable insights to guide developers towards energy efficiency for future versions of their software. 

A deep investigation of these causes would involve computing the source-level diff between two versions, which is then analyzed line by line for changes that might affect energy consumption. This is not only tedious because the code bases of the JVM and Spring Boot (including their dependencies) comprise millions of lines of code, and it may not always be clear to what extent a change in the code has an actual effect on energy consumption.
Therefore, we suggest a heuristic approach to identifying energy-relevant changes based on an analysis of \emph{release notes}, which we want to explore further in the future.
This approach consists of three steps: 1) the release notes of a given version (of the JVM or Spring Boot and their dependencies) are analyzed for any changes from the previous version that typically influence energy consumption; 2) the changes identified in the first step are individually measured with micro-benchmarks and, if shown to be significant, 3) the results of the micro-benchmarks are discussed for relevance.

While such an approach should be practically feasible, it builds on many assumptions that must be carefully documented and considered when such an approach is evaluated in the future. While the analysis of release notes seems to be a promising research endeavor~\citep{bi_empirical_2022}, it must be noted that only around 26\,\% of all changes are documented in the release notes~\citep{abebe_empirical_2016}.
Therefore, we suggest that important software components, such as the JVM or Spring Boot, should be extended by their developers with an energy benchmark that measures the energy consumption of the most relevant mechanisms, such as the Just-in-Time (JIT) compiler, garbage collection, or memory management (for the JVM) and dependency injection, reflection, and I/O libraries (for Spring Boot). Such an energy benchmark can be executed for every release, and the results published as part of the release notes.

\section*{Acknowledgements}\label{acknowledgements}
We thank our colleagues and hardware experts Andreas Breitenmoser, Hans Dermot Doran, and Carlos Rafael Tordoya Taquichiri for the valuable discussions on the energy-specific differences of the hardware platforms used in the case study. We thank our colleague Marcela Ruiz for valuable feedback on an earlier version of this paper.

\section*{Declaration of competing interest}
The authors declare that they have no known competing financial interests or personal relationships that could have appeared to influence the work reported in this paper.

\section*{Declaration of generative AI and AI-assisted technologies in the writing process}
During the preparation of this work, the authors used DeepL and Copilot to check and improve the grammar of the manuscript and to generate the initial versions of some of the diagramming code and summaries. Copilot was also used to help write the benchmark scripts and the Python code for the statistical analysis. Having used these tools, the authors reviewed and edited the content and code as necessary, taking full responsibility for the published article.

\section*{Data availability}
The datasets generated and analyzed in the context of this study, the raw data and the detailed plots, as well as additional resources that are useful for reproducing our research, are available \citep{ZenodoDataset:2025}. Around 10 kWh of electricity was used to run the benchmarks to produce the dataset. The authors have compensated for that amount via myclimate.org. 

\bibliographystyle{elsarticle-harv} 
\bibliography{cas-refs,bibliography}

\end{document}